\documentclass[reprint,aps,amsmath,pra,showpacslongbibliography]{revtex4-2}
\usepackage{graphicx}
\usepackage[ colorlinks = true,
       linkcolor = blue,
       urlcolor = blue,
       citecolor = red,
       anchorcolor = green,
]{hyperref}\usepackage{graphicx}
\usepackage{bbold,soul,amsmath,amssymb}
\usepackage{xcolor}

\usepackage{float}

\newcommand{\norm}[1]{\left\lVert#1\right\rVert}

\usepackage{tikz}
\usetikzlibrary{quantikz2}
\usetikzlibrary{matrix}
\usepackage{algorithm}
\usepackage{algpseudocode}
\begin{document}
\tikzset{meter/.append style={draw, inner sep=10, rectangle, font=\vphantom{A}, minimum width=30, line width=.8,
 path picture={\draw[black] ([shift={(.1,.3)}]path picture bounding box.south west) to[bend left=50] ([shift={(-.1,.3)}]path picture bounding box.south east);\draw[black,-latex] ([shift={(0,.1)}]path picture bounding box.south) -- ([shift={(.3,-.1)}]path picture bounding box.north);}}}

\title{Characterizing Gaussian quantum processes with Gaussian resources}

\email{kevin.valsonjacob@wheaton.edu}
\author{Logan W. Grove\textsuperscript{1}}
\author{Pratik J. Barge\textsuperscript{2}}
\author{Kevin Valson Jacob\textsuperscript{1}}

\affiliation{\textsuperscript{1}Department of Physics and Engineering, Wheaton College, Wheaton, Illinois 60187, USA}
\affiliation{\textsuperscript{2}Department of Physics, Washington University in St Louis, Missouri, 63105, USA
}

\date{\today}

\begin{abstract}

Characterizing quantum processes is indispensable for the implementation of any task in quantum information processing. In this paper, we develop an efficient method to fully characterize arbitrary Gaussian processes in continuous-variable quantum systems. This is done by directly obtaining all elements of the symplectic matrix that describes the process. Only Gaussian resources such as coherent probes and quadrature measurements are needed for this task. The method is efficient, involving only $O(N^2)$ steps to characterize an $N$-mode system. Further, the method is resilient to uniform loss. We simulate this procedure using the Python package Strawberry Fields. We observe that heterodyne measurements outperform homodyne measurements for reconstructing Gaussian processes.

\end{abstract}

\maketitle

\section{Introduction}
A variety of tasks in quantum information science can be implemented by continuous variable systems, of which quantum light is a leading platform~\cite{silberhorn2010,braunstein2005}. Such implementations necessitate the production, transformation, and detection of quantum light. Recently, several platforms have been developed for these tasks~\cite{Raffaelli_2018, Madsen2022}. Due to manufacturing imperfections, a post-production characterization of these technologies is necessary to ensure that such devices function as intended~\cite{Rhode_review}. 

 Characterizing continuous variable systems can be inherently challenging, as the Hilbert space corresponding to such systems is infinite-dimensional~\cite{weedbrook:12}. However, a simplification is possible when we restrict our attention to Gaussian systems. These systems include linear optical systems as well as those with quadratic nonlinearities such as parametric amplifiers and down-converters, and squeezers~\cite{caves,braunstein_squeezing}. When used in conjunction with non-Gaussian systems, Gaussian systems enable the execution of several important quantum information tasks including communication, cryptography, teleportation, and non-universal quantum computation such as Gaussian boson sampling~\cite{weedbrook:12}.

 Gaussian evolutions are generated by Hamiltonians that are at most quadratic in the mode operators. These evolutions map Gaussian states \textemdash described by their first and second moments alone \textemdash to Gaussian states. Therefore, characterizing an $N$-mode Gaussian evolution necessitates the finding of $O(N^2)$ parameters.

A general scheme to characterize quantum devices consists of three steps. First, well-characterized quantum states are prepared. Subsequently, these states are used to probe the uncharacterized device. Finally, a specified set of measurements is performed on the states after they have evolved through the device. From the outcomes of these measurements, one then deduces the complete description of the unknown device.

Several investigations have focused on characterizing linear optical systems by determining their $N\times N$ unitary transfer matrix. This was done using non-Gaussian probes in Refs.~\cite{peruzzo2011multimode, Obrien2012Unpub, Dhand_2016, Zhou:15} and with Gaussian probes in Refs.~\cite{Rahimi-Keshari:13, Laing_Phase_Lift, Hoch_2023}. However, these schemes assumed that the unitary matrix is real bordered, i.e. the elements in its first row and first column are real. This restriction was removed in Ref.~\cite{valson2018}, which considered not only linear optical systems but also those with quadratic nonlinearities i.e. all Gaussian systems. However, nonclassical probes were still used in this analysis. 

Several other works have explored beyond linear optics by characterizing all Gaussian systems. In Ref.~\cite{Wang_2013} the Husimi $\it{Q}$ function corresponding to a Gaussian process was found using coherent probes. In contrast, coherent probes and photon number-resolving measurements were used to directly find the parameters of the Gaussian map in Ref.~\cite{Kumar2020}. Reference ~\cite{Teo_2021} uses coherent states and heterodyne measurements to characterize a single-mode Gaussian process.

Despite this progress, a method that directly characterizes multimode Gaussian processes with only Gaussian resources is still needed. In this paper, we develop such a method, using no nonclassical resources, and employing only coherent probes and quadrature detection. Using this method, we directly obtain all elements of the symplectic matrix that completely describes the process. 

This paper is structured as follows: In Sec.~\ref{sec:preliminaries} we outline the mathematical conventions used in this work. Then in Sec.~\ref{sec:setup} we give details of our proposed setup. Section~\ref{sec:Characterization} then shows the procedure employed in our method. In section~\ref{sec:loss}, we consider experimental limitations that can constrain implementations of the proposed method. The results of simulating our method are presented in Sec.~\ref{sec:SF}. We discuss salient results in Sec.~\ref{sec:discussion}, and conclude in Sec.~\ref{sec:conclusion}.

\section{Quantum Continuous Variable Systems}
\label{sec:preliminaries}

Consider an $N$-mode bosonic quantum system that has the corresponding mode annihilation and creation operators $\{\hat{a}_i,\hat{a}_i^\dagger\}_{i=1}^N$. These operators satisfy the commutation relation $\left[\hat{a}_i,\hat{a}_j^\dagger\right] = \delta_{ij}$.
These operators can be used to define a set of quadrature operators for each mode as
\begin{equation}
\hat{X}_i = \frac{1}{\sqrt{2}}\left(\hat{a}_i + \hat{a}_i^\dagger\right), \qquad
\hat{P}_i = \frac{1}{\sqrt{2}i}\left(\hat{a}_i - \hat{a}_i^\dagger\right), 
\label{eq:quad_definition}
\end{equation}
where the $\hat{X}$ quadratures are called position operators, and the $\hat{P}$ quadratures are called the momentum operators. 
These quadrature operators are Hermitian and as such have real eigenvalues which can be experimentally measured. 

As there are $2N$ of these quadrature operators, we formally arrange these quadrature operators as a vector of $2N$ elements as
\begin{align}
  {\bf{\hat{r}}} \equiv \left(\hat{X}_1,\dots,\hat{X}_N,\hat{P}_1,\dots,\hat{P}_N\right)^\mathrm{T}.
\end{align}
\subsection{Gaussian states}
A Gaussian state $\hat{\rho}$ of a continuous variable system is fully characterized by the first and second moments of the vector of quadrature operators. The first moment is called the mean vector and is defined as 
\begin{equation}
  {\bf{r}} \equiv \left\langle {\bf{\hat{r}}}\right \rangle = \mathrm{Tr}\left({\bf{\hat{r}}}\hat{\rho}\right).
\end{equation}
The second moment, called the covariance matrix, is defined as
\begin{equation}
  {\bf{\sigma}} = \mathrm{Tr}\left[\left\{ ({\bf{\hat{r}}}-{\bf{r}}),({\bf{\hat{r}}}-{\bf{r}})^\mathrm{T}\right\} \hat{\rho}\right],
\end{equation}
where $\{.,.\}$ represents the anticommutator. Examples of Gaussian states are the vacuum state, coherent states, thermal states, and squeezed states~\cite{wang2007}.
\subsection{Gaussian processes}
Gaussian quantum processes map Gaussian states to Gaussian states. Such processes include both passive transformations of the mode operators, as in the case of linear optics, and active transformations that do not preserve energy. We consider Gaussian processes that map the mean vector of a Gaussian state as
\begin{equation}
  {\bf{r}}^{\text{out}} = S{\bf{r}}^{\text{in}}, 
  \label{eq:Gaussian_map}
\end{equation}
where $S$ is a $2N\times2N$ invertible symplectic matrix of real elements with determinant $+1$~\cite{weedbrook:12,Adesso_2014,serafini2017quantum}. Since $S$ is symplectic, we have 
\begin{equation}
  SJ S^T = J, \qquad 
  J = \begin{pmatrix}
    0_N & \mathbb{1}_N \\
    -\mathbb{1}_N & 0_N
  \end{pmatrix}.
\end{equation} 
\subsubsection*{Passive transformations}

If the Gaussian quantum process corresponds to passive transformations of the mode operators, as in linear optics, the $2N\times2N$ symplectic matrix $S$ is related to a $N\times N$ unitary matrix $U$ that transforms the mode operators of the linear optical system~\cite{Valson2020}. It is shown in Appendix \ref{appendix:Relate_U_to_S} that $U$ is related to $S$ as
\begin{equation}
  S_{\text{passive}} = \begin{pmatrix}
     \mathbb{Re}(U) & \mathbb{Im}(U) \\
     -\mathbb{Im}(U) & \mathbb{Re}(U)
   \end{pmatrix}.
   \label{eq:S_U_relation}
\end{equation}

\subsection{Gaussian measurements}
A quantum measurement is defined to be a Gaussian measurement if its application to a Gaussian state yields a Gaussian-distributed outcome~\cite{weedbrook:12}. Two common Gaussian measurements are the homodyne and the heterodyne. Both of these are single-mode measurements that can be used to reconstruct the first moment (i.e. the mean vector) of a quantum state.

Homodyne detection is modeled as a projection onto the quadrature basis~\cite{weedbrook:12}. As such, it can measure any given quadrature of a quantum state at a time. In contrast, heterodyne detection is modeled as a projection onto the coherent basis. Thus, it jointly measures complementary quadratures, and is subject to additional Arthurs-Kelly type noise that is inherent to such joint measurements~\cite{Teo2017HomoHetero}.

\section{Setup}
\label{sec:setup}
\begin{figure}[H]
\resizebox{0.5\textwidth}{!}{
  \centering
  \
    \includegraphics[width=\columnwidth]{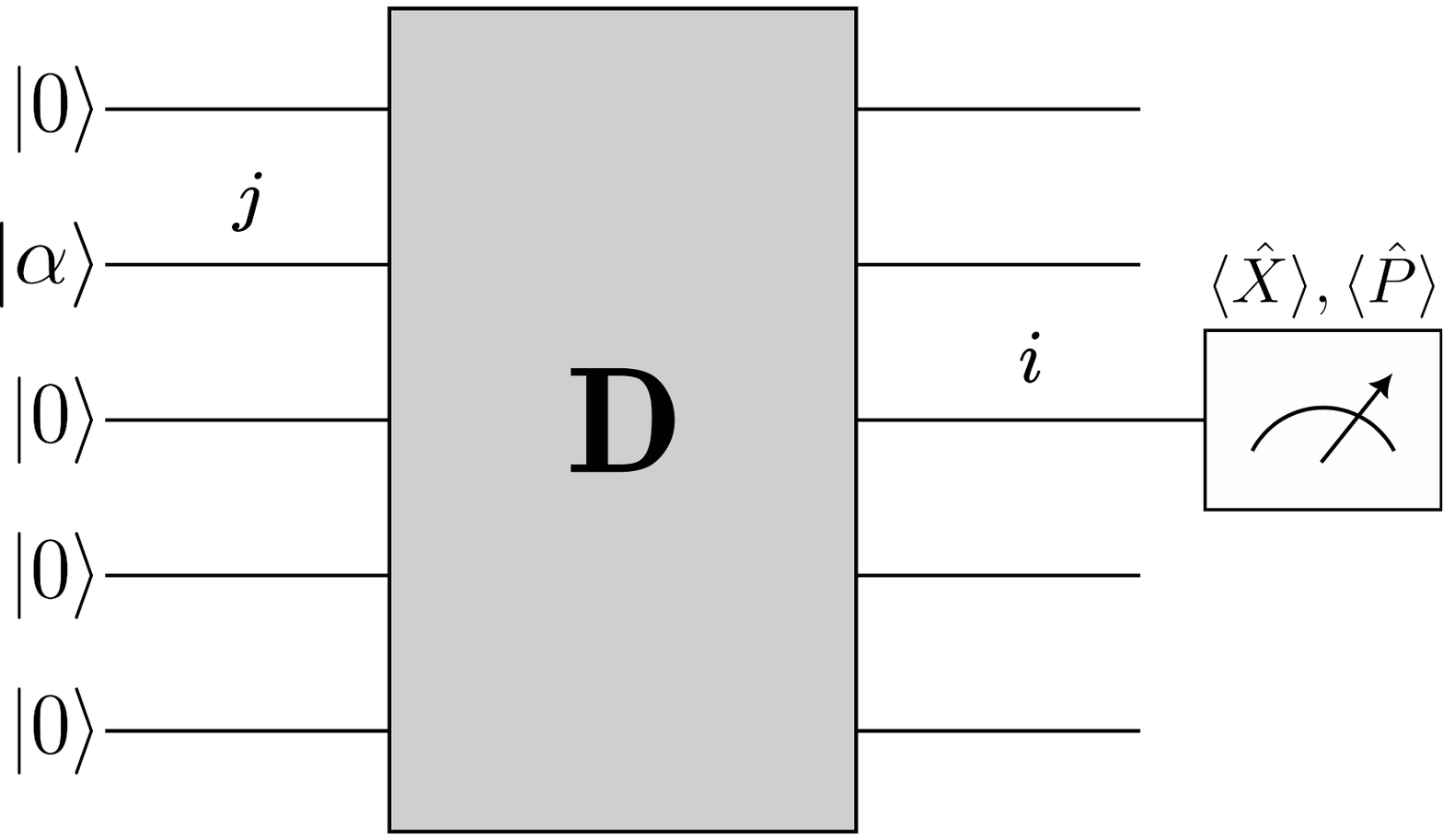}
  }
  \caption{A schematic for characterizing Gaussian devices. A coherent probe $|\alpha\rangle$ is input into the $\it{j}{\text{th}}$ mode of a device $D$. After the probe evolves through the device, output quadratures are measured in each output mode.}
  \label{fig:setups}
\end{figure}

A schematic of the setup proposed to characterize the Gaussian process is shown in Fig.~\ref{fig:setups}.
As shown in Fig.~\ref{fig:setups}, the setup  consists of a Gaussian device $D$ with $N$ input and output modes. A well-characterized coherent probe is sent into the $\it{j}{\text{th}}$ input mode of the device and vacuum states are input into every other mode. The coherent probe then transforms through the device following the map of Eq.~\eqref{eq:Gaussian_map}. Quadratures are measured at the $\it{i}{\text{th}}$ output mode, either by homodyne detection or by heterodyne detection. This process is repeated for all combinations of $i,j\in \{1,2,\ldots, N\}$.

\section{Characterization of Gaussian processes}
\label{sec:Characterization}
We now outline our method for characterizing Gaussian processes. This involves finding all $4N^2$ real elements of $S$. 

When a coherent probe is input to the $j^{\text{th}}$ input mode and vacuum states input into every other mode, the input mean vector is 
\begin{equation}
{\bf{r}}^{\text{in}} = \left( 0, \; \cdots, \; \langle \hat{X}_j \rangle^{\text{in}}, \; 0, \; \cdots, \; \langle \hat{P}_j\rangle ^{\text{in}} , \; \cdots, \; 0 \right)^{\text{T}}.
\end{equation} 
This state then evolves via the linear map of Eq.~\eqref{eq:Gaussian_map} representing the device.
Subsequently, the output mean vector ${\bf{r}}^{\text{out}}$ will be

\begin{equation}
   {\bf{r}}^{\text{out}} = \langle\hat{X}_j\rangle^{\text{in}}\left( \begin{array}{c} S_{1, j} \\ S_{2, j} \\ \vdots \\ S_{2N, j} \end{array} \right) 
    + \langle\hat{P}_j \rangle^{\text{in}} \left( \begin{array}{c} S_{1, N+j}  \\ S_{2, N+j} \\ \vdots \\ S_{2N, N+j}  \end{array} \right).
    \label{eq:r_out}
\end{equation}

By controlling the phase of the input coherent state, we may choose either $\langle \hat{X}_j \rangle^{\text{in}}$ or $\langle \hat{P}_j \rangle^{\text{in}}$ to be zero independently of the other. In both of these cases, each measurement of an output quadrature is related to a single element of $S$. 

For instance, in order to find the element $S_{i,j}$ where $i,j\in\{1,2,\ldots, N\}$, we send in a coherent state $|\alpha\rangle$ where $\alpha \in \mathbb{R}$ to the input mode $j$. Subsequently, we measure the position quadrature in mode $i$, i.e. $\langle \hat{X}_i\rangle^{\text{out}}$. Following Eq.~\eqref{eq:r_out}, and noting that $\langle \hat{P}_j\rangle^{\text{in}} = 0$, this measurement yields
\begin{align}
  \langle\hat{X}_i\rangle^{\text{out}} &= \langle\hat{X}_j\rangle^{\text{in}} S_{i,j} \nonumber \\
  &= \sqrt{2}\alpha S_{i,j}.
\end{align}
This relation enables one to directly obtain the matrix element $S_{i,j}$ as
\begin{equation}
  S_{i,j} = \frac{\langle\hat{X}_i\rangle^{\text{out}}}{\sqrt{2}\alpha}.
  \label{eq:matrix_element_1/alpha}
\end{equation}

Likewise, we obtain $S_{N+i,j}$ where $i,j\in\{1,2,\ldots, N\}$ by measuring the momentum quadrature instead, i.e. $\langle \hat{P}_i\rangle^{\text{out}}$. 

A similar process yields the elements $S_{i,N+j}$ and $S_{N+i,N+j}$ as well.
To do this, we probe the device with an input coherent state $|i\alpha\rangle$ where $\alpha \in \mathbb{R}$, i.e. $\langle \hat{X}_j \rangle^{\text{in}} = 0$. Measuring the position and momentum quadratures at the output then yields these elements.

\begin{figure}[H]
  \centering
  \includegraphics[width=\columnwidth]{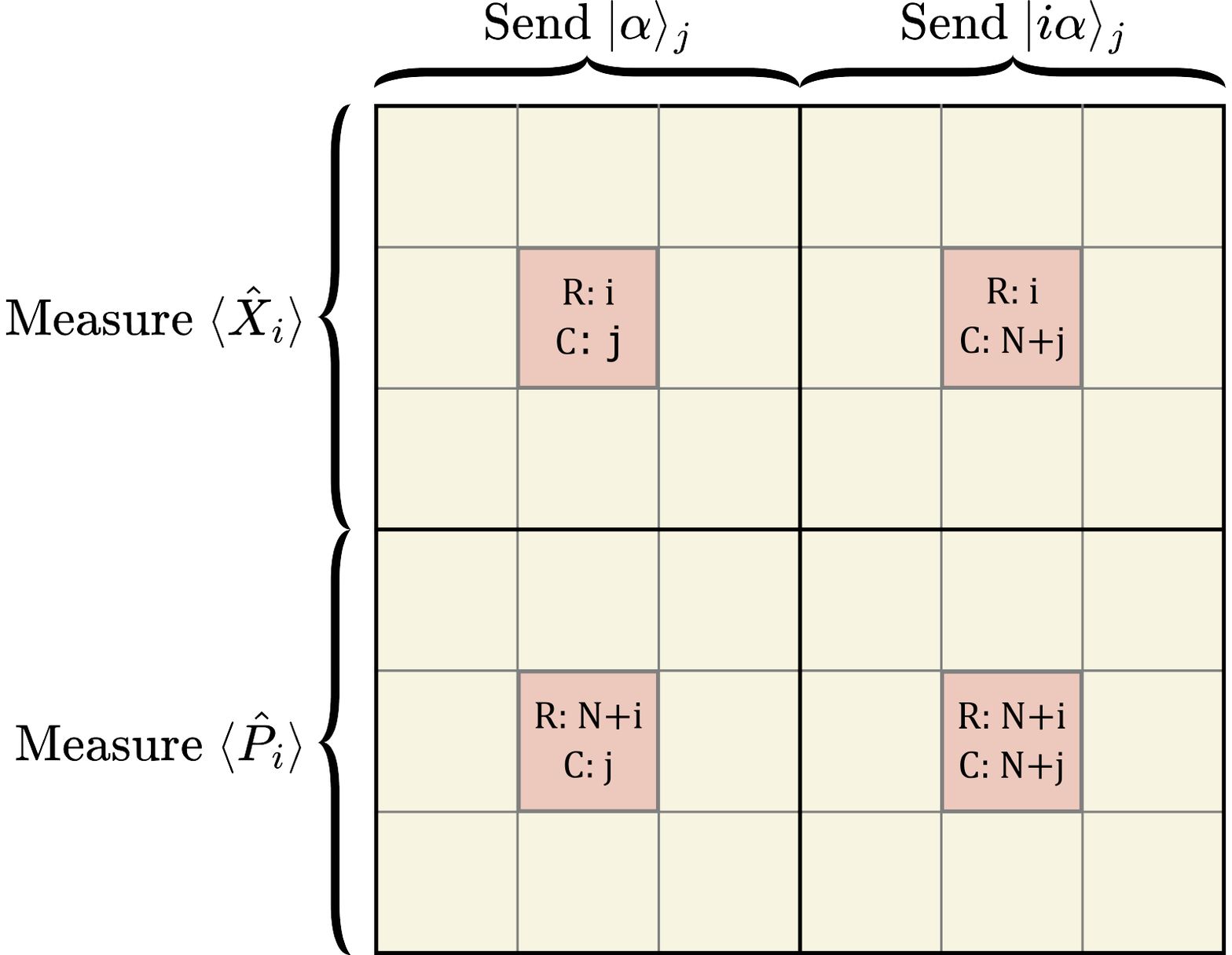}
  \caption{ The probes and measurement settings needed to find the elements of the matrix $S$. $\text{R}$ denotes row number and $\text{C}$ denotes column number. Four elements of the matrix can be determined for a given input and output mode by choosing the phase of the probe as well as the quadrature measured.}
  \label{fig:S_element}
\end{figure}

Finally, we determine all elements of $S$ by choosing all combinations of input modes $j$ and output modes $i$. 
Figure ~\ref{fig:S_element} shows the measurement settings necessary to find all elements of $S$, requiring $4N^2$ measurement settings in total.

\subsection*{Simplification for passive transformations}
Passive transformations, such as those in linear optics, can be characterized by $O(N^2)$ parameters in the form of a $N\times N$ unitary matrix $U$. 
Noticing the relation of $U$ to $S$ in Eq.~\eqref{eq:S_U_relation}, we can reduce the number of measurements required to completely characterize $U$. For instance, by restricting the input coherent states to be of the form $|\alpha\rangle$ where $\alpha \in \mathbb{R}$, one can find all the elements of $\mathbb{Re}(U)$ and $\mathbb{Im}(U)$. The unitary matrix can then be reconstructed as $\mathbb{Re}(U)+i\mathbb{Im}(U)$. Thus, only $2N^2$ measurement settings are needed to characterize an $N$-mode linear optical device.

\section{Modeling experimental constraints}
\label{sec:loss}
In this section, we account for two realistic limitations that can constrain implementations of the proposed method: optical loss and phase-modulation errors.

\subsection{Optical loss}
\begin{figure}[H]
  \centering
  \includegraphics[width=0.65\columnwidth]{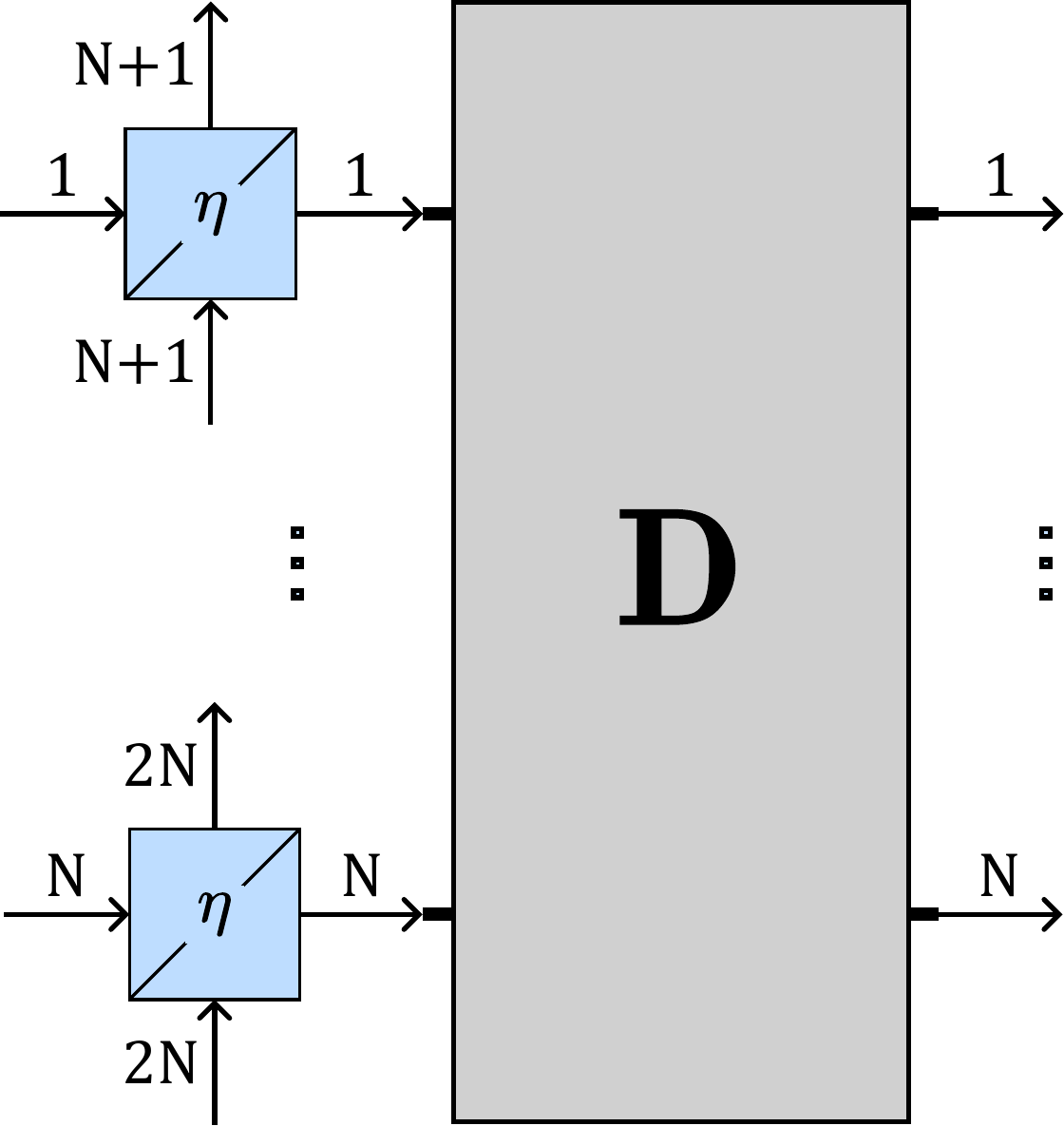}
  \caption{Balanced loss model with fictitious beam splitters of the same transmissivity $\eta$ at the input. For each input mode $j$, we label the ancillary mode of the fictitious beam splitter as $N+j$. The ancillary modes $\{N+1, \ldots, 2N\}$ have vacuum as input.}
  \label{fig:loss}
\end{figure}

Realistically, optical loss is inevitable to any experiment. Such loss may occur while light is input into the components of the device (input loss) as well as while light propagates through the device (propagation loss)~\cite{Chanana2022}. To model these, we consider a uniform (or balanced) loss model, where loss is the same in each mode. Such loss is modeled by a set of fictitious beam splitters of the same transmissivity $\eta \in (0,1]$ attached to each input mode of the device~\cite{Gerry_Knight_2023}. Vacuum is input into all of the ancillary modes. A schematic of this loss model is shown in Fig.~\ref{fig:loss}.

It was shown in Ref.~\cite{Oszmaniec2018} that uniform loss commutes with passive linear optical transformations. As shown in Appendix \ref{appendix:loss_commutes}, uniform loss commutes with the Gaussian map of Eq.~\eqref{eq:Gaussian_map} also. Therefore, this loss model is agnostic to the position of loss in any device with uniform loss. This allows us to transfer any uniform loss within the device to its input. Thus, the effect of uniform loss is to attenuate the input coherent probe $|\alpha \rangle_j$ to $|\sqrt{\eta}\,\alpha \rangle_j$. Consequently, Eq.~\eqref{eq:r_out} is modified to include the loss as 
\begin{equation}
   {\bf{r}}^{\text{out}} = \sqrt{\eta}\langle\hat{X}_j\rangle^{\text{in}}\left( \begin{array}{c} S_{1, j} \\ S_{2, j} \\ \vdots \\ S_{2N, j} \end{array} \right) 
    + \sqrt{\eta}\langle\hat{P}_j \rangle^{\text{in}} \left( \begin{array}{c} S_{1, N+j}  \\ S_{2, N+j} \\ \vdots \\ S_{2N, N+j}  \end{array} \right).    \label{eq:r_out_loss}
\end{equation}
Therefore, following the procedure of section \ref{sec:Characterization} yields an experimentally obtained matrix $\tilde{S}$. $\tilde{S}$ is related to the true symplectic matrix $S$ by $\tilde{S} = \sqrt{\eta}S$. By considering that $\det(S) = 1$, we obtain the transmissivity $\eta$ as

\begin{equation}
  \det(\tilde{S}) = (\sqrt{\eta})^{2N} \det(S) = \eta^{N}.
\end{equation}

Thus, $\eta$ is directly determined from experimental data, thereby requiring no additional measurements. Once determined, $\eta$ is used to recover $S$ from the experimentally obtained matrix $\tilde{S}$ as $S = \frac{1}{\sqrt{\eta}}\tilde{S}$. This procedure is summarized in Algorithm~\ref{alg:measurequadratures}.

A similar method to characterize loss can be employed for passive linear optical transformations, as the determinant of a unitary transformation also has unit modulus.

\begin{algorithm}[H]
\caption{Algorithm for reconstructing $S$}
\label{alg:measurequadratures}
\begin{algorithmic}[1]
\Procedure{ReconstructGaussianProcess}{}
  
   \For{$j$ $\in \{1, \dots, N\}$ and $\alpha \in \mathbb{R}$ }
     \State Probe $j^{\text{th}}$ mode with $|\alpha\rangle$ 
     \State Evolve $|\alpha\rangle$ through the unknown device
     \State Measure $\langle \hat{X} \rangle$ and $\langle \hat{P} \rangle$ over all output modes 
     \State Reconstruct $j^{\text{th}}$ column of $\tilde{S}$ 
      
    \State Probe the $j^{\text{th}}$ mode with $|i\alpha\rangle$ 
    \State Evolve $|i\alpha\rangle$ through the unknown device      
     \State Measure $\langle \hat{X} \rangle$ and $\langle \hat{P} \rangle$ over all output modes 
    \State Reconstruct the $(j+N)^{\text{th}}$ column of $\tilde{S}$
  \EndFor
  
  \State Calculate $\eta = \det(\tilde{S})^{\frac{1}{N}}$
  \State Calculate $S = \frac{1}{\sqrt{\eta}}\tilde{S}$  
\EndProcedure
\end{algorithmic}
\end{algorithm}
\subsection{Phase-modulation errors}
As the method proposed requires the precise modulation of the phases of the input probes, we investigate the effect of  phase-modulation errors. To this end, without loss of generality, we consider the reconstruction of a symplectic matrix element $S_{i,j}$ where $i,j\in\{1,2,\ldots, N\}$. The probe used is labeled $|\alpha = re^{i\phi}\rangle$, where $r \in \mathbb{R}$ and $\phi$ models the phase-modulation error. From Eq.~\eqref{eq:r_out}, the measured output quadrature is 
\begin{align}
      \langle\hat{X}_i\rangle^{\text{out}} &= \langle\hat{X}_j\rangle^{\text{in}} S_{i,j} + \langle\hat{P}_j\rangle^{\text{in}} S_{i,N+j}\nonumber \\
      &=\sqrt{2}r\cos{\phi}\,S_{i,j} + \sqrt{2}r\sin{\phi}\,S_{i,N+j}\nonumber \\
      &= \sqrt{2}r S_{i,j} + \sqrt{2}r \phi S_{i,N+j} + O( \phi^2).
\end{align}
This relation enables us to find $S_{i,j}$ as 
\begin{equation}
    S_{i,j} = \frac{\langle\hat{X}_i\rangle^{\text{out}}}{\sqrt{2}r} - \phi  S_{i,N+j} + O(\phi^2).
    \label{eq:matrix_element_phase_error}
\end{equation}
Comparing Eq.~\eqref{eq:matrix_element_phase_error} with Eq.~\eqref{eq:matrix_element_1/alpha}, we note that phase-modulation error $\phi$ introduces a first-order error in the reconstructed matrix element. If we further assume $\phi$ to be randomly distributed with a mean value of zero, then averaging our results over many trials suppresses the reconstruction error to  second order in $\phi$. Thus the reconstruction is robust to small phase-modulation errors.

%\subsection{Scalability with the coherent probe intensity}
%\label{sec:intensity_scalability}
%In the proposed method, we obtain the elements of the unknown symplectic matrix as a ratio of the output and input quadratures. Therefore, we expect that the input probe intensity will affect the accuracy with which we determine these matrix elements.  From Eq.~\eqref{eq:matrix_element_1/alpha}, we can obtain the uncertainty with which the matrix elements are recovered as 
%\begin{equation}
%  \left|\Delta S_{i,j}\right| \propto\frac{1}{|\alpha|^2}.
%  \label{eq:matrix_element_uncertainty}
%\end{equation}
%Thus a larger probe intensity shall result in a greater accuracy with which the matrix elements are determined. 

\section{Simulation}
\label{sec:SF}
To test the validity of our method, we simulate it on Strawberry Fields, an open-source Python library for quantum continuous variable systems \cite{Killoran2019strawberryfields, Bromley_2020}. The corresponding Python code used for the simulation is available on GitHub \cite{Grove2025}. For each simulation, the accuracy of reconstruction was quantified by a scaled Frobenius (Hilbert–Schmidt) norm of the difference between a randomly generated symplectic matrix $S^\text{rand}$ and its reconstruction $S^\text{recon}$. This norm is defined as
\begin{align}
  F&=\frac{1}{N}\norm{S^\text{rand}-S^\text{recon}}_F \nonumber \\ &= \frac{1}{N}\sqrt{\sum_{i,j}^{2N}\left|S^\text{rand}_{i,j}-S^\text{recon}_{i,j}\right|^2},
\end{align} 
where $N$ is the number of modes. A lower value of $F$ indicates a more accurate reconstruction.

\subsection{Scalability with the number of modes}
\label{sec:Scalability with the number of modes}
To demonstrate the scalability of the reconstruction with the number of modes, we randomly generated symplectic matrices $S^\text{rand}$ for $N$-mode devices where $N$ varied from 2 to 20. For each case, we implemented the corresponding Gaussian transformation as well as uniform loss. We then reconstructed those devices using either homodyne or heterodyne detection to obtain $S^\text{recon}$. Finally, we calculated the accuracy of reconstruction $F$.

To eliminate any bias in our initial choice of $S^\text{rand}$, we repeated this process 500 times and calculated the average value of $F$ for each choice of $N$. 
In our simulation, we used coherent probes of amplitude $\alpha = 1000$. The transmissivity $\eta$ was set at either 0\% or at 50\%. We ensured that the same number of coherent probes were used in both homodyne and heterodyne reconstruction. This was necessary because homodyne reconstruction requires two probes to measure both quadratures while heterodyne only requires one. 

\begin{figure}[H] 
\centering
\includegraphics[width=1.1\columnwidth]{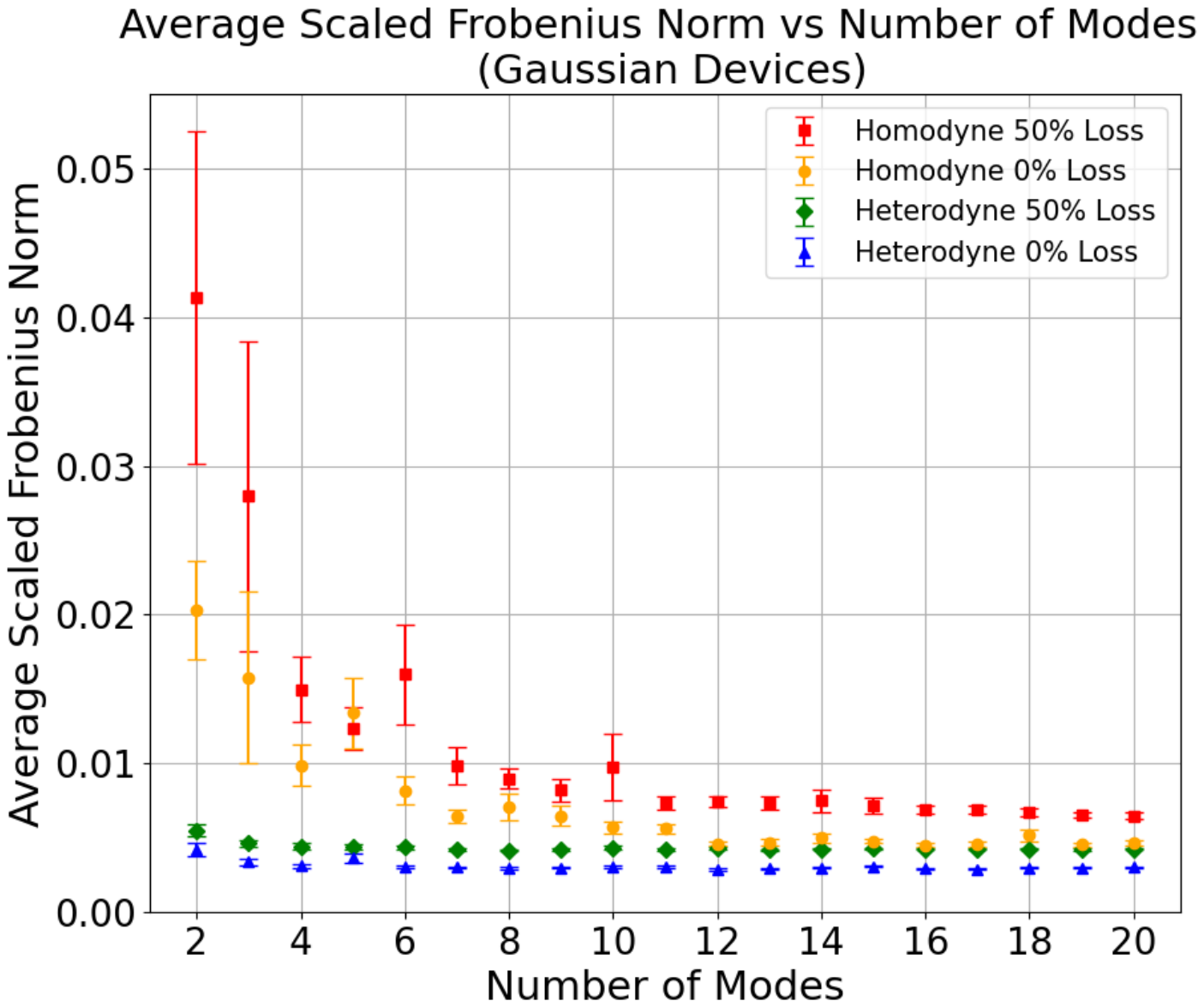}
\caption{Accuracy of reconstruction as a function of the number of modes.  Reconstruction was performed using both heterodyne and homodyne measurements. The loss parameter was set either at 0\% or at 50\%. Heterodyne reconstruction is observed to outperform homodyne reconstruction.} 
\label{fig:gaussian_norm}
\end{figure}

The results of the simulation are shown in Fig.~\ref{fig:gaussian_norm}. We note that $F$ tends to be independent of the number of modes of the device. This indicates that our method is readily scalable with the size of the reconstructed device. We also notice that heterodyne reconstruction outperforms homodyne reconstruction. This is supported by two observations. First, for each value of the loss parameter, heterodyne reconstruction yields a lower value of the scaled Frobenius norm than that of homodyne reconstruction. Second, the uncertainties in the homodyne data are larger than those in heterodyne data. 

\begin{figure}[ht!]
\centering
\includegraphics[width=1.2\linewidth]{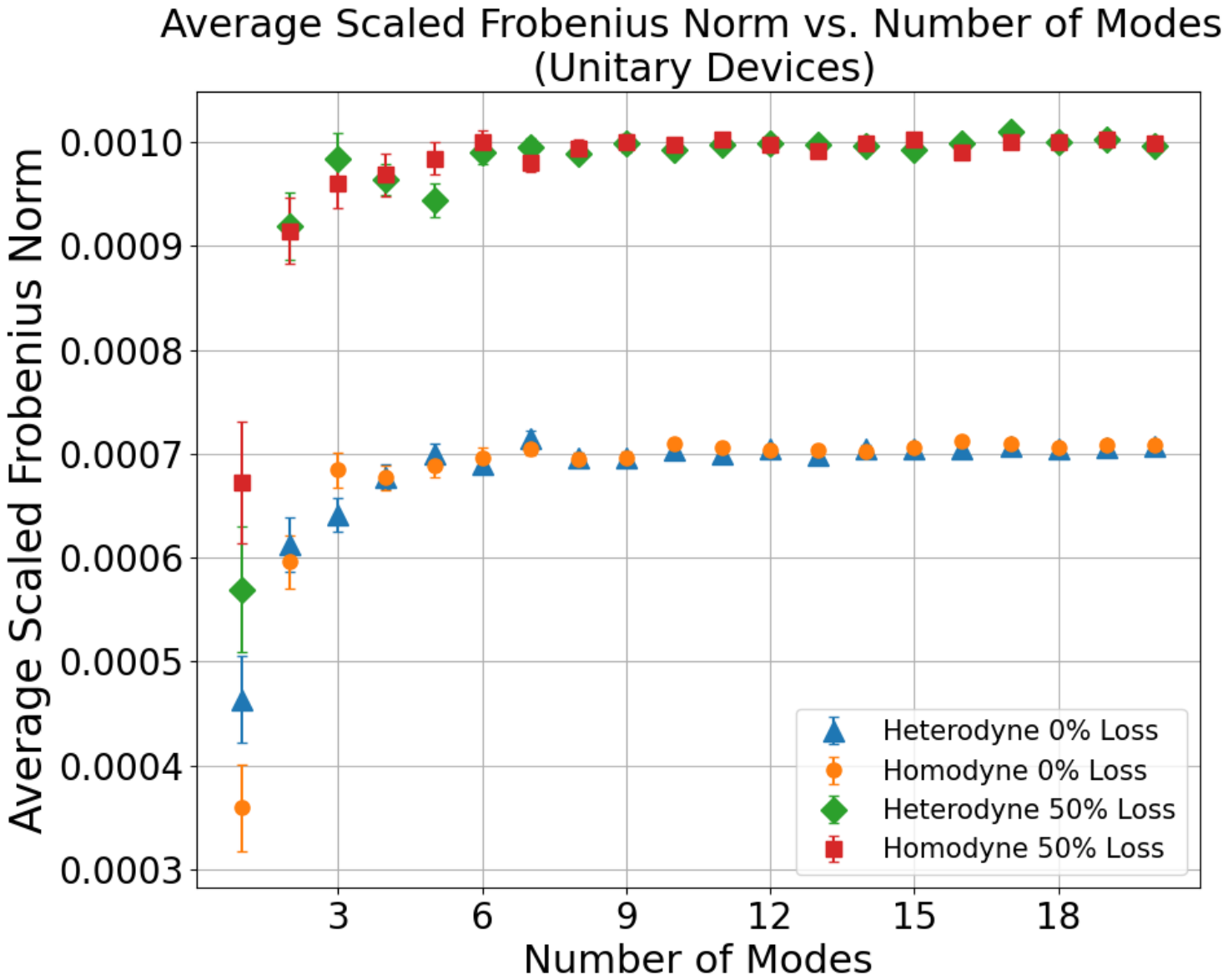}
\caption{Accuracy of reconstruction as a function of the number of modes.  Reconstruction was performed using both heterodyne and homodyne measurements for lossless and lossy linear devices. The accuracy is observed to be similar for both homodyne and heterodyne reconstruction.}
\label{fig:unitary_norm}
\end{figure}

A similar simulation was performed for linear processes. In this simulation, unitary matrices were randomly sampled from the Haar measure. After reconstruction, the scaled Frobenius norm of the difference between the random unitary and the reconstructed unitary was computed. This task was repeated 50 times and the average scaled Frobenius norm was calculated. 

 The results of this simulation are shown in Fig.~\ref{fig:unitary_norm}. As in the case of Gaussian devices, the accuracy of reconstruction is observed to be independent of the number of modes of the device, thereby indicating the scalability of the procedure. However, we observe that both homodyne detection and heterodyne detection reconstruct linear devices equally well. 
\subsection{Scalability with the coherent probe intensity}
In the proposed method, we obtain the elements of the unknown symplectic matrix as a ratio of the output and input quadratures, as seen in Eq.~\eqref{eq:matrix_element_1/alpha}. Therefore,  the input probe intensity will affect the accuracy with which we determine the unknown symplectic matrix elements. A larger probe intensity shall increase the accuracy of reconstructing individual matrix elements, as can be noted from Eq.~\eqref{eq:matrix_element_1/alpha}. This limit in the accuracy of reconstruction follows the standard quantum limit as we use only classical resources in the reconstruction \cite{lee2002rosetta}.

As experimental constraints may restrict the intensity of coherent probes used, one may wonder if the reconstruction method is accurate only with sufficiently strong coherent probes. We answer this question by demonstrating that the reconstruction method works for arbitrarily weak coherent probes. While using weak coherent probes, we  repeat the method over multiple trials and average over the recovered matrix element. This  repetition alone can reduce the standard error in measurements that scales inversely with $\sqrt{n}$ where $n$ is the number of trials.

\begin{figure}[ht!]
    \centering
    \includegraphics[width=1.1\linewidth]{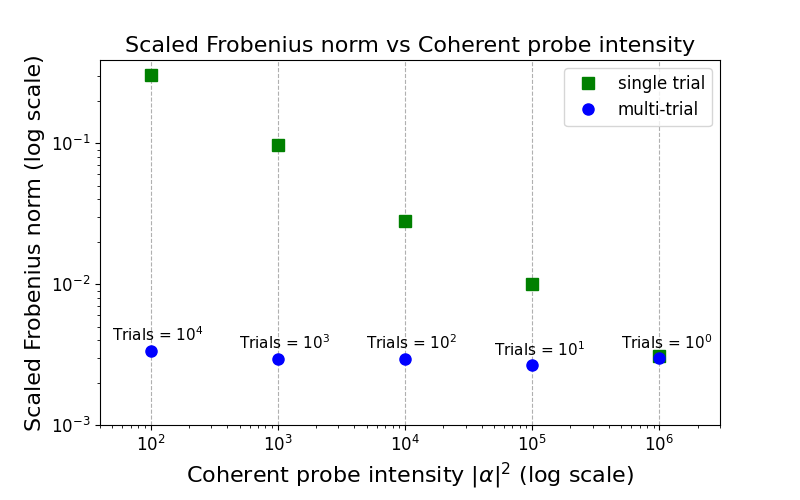}
    \caption{Accuracy of reconstruction as a function of the the intensity of the coherent probes used. A 5-mode Gaussian device was randomly generated and then reconstructed using heterodyne measurements. For each coherent probe intensity, reconstruction was performed using a single trial as well as with a variable number of trials $n$. The accuracy of reconstruction is the same in cases where $n|\alpha|^2$ is the same.}
    \label{fig:intensity_scaling}
\end{figure}

To demonstrate this numerically, we randomly generated the symplectic matrix corresponding to a five-mode Gaussian device. We reconstructed it using heterodyne measurements for a range of coherent probe intensities. The reconstruction was performed using a single trial as well as with a multiple number of trials. The accuracy of reconstruction was calculated for each case, and plotted against the coherent probe intensity in Fig.~\ref{fig:intensity_scaling}. 

In Fig.~\ref{fig:intensity_scaling}, we note that the accuracy of reconstruction is the same for multiple coherent probe intensities provided that reconstruction with weaker probes was repeated multiple times. This is supported by the observation that the scaled Frobenius norm is the same as long as $n|\alpha|^2$ (or $\sqrt{n}|\alpha|$) remains the same. Since $n|\alpha|^2$ is a measure of the average number of photons used to reconstruct any given matrix element, the accuracy of reconstruction is limited by the energy of the probes alone. Therefore one may use weaker coherent probes to obtain the same reconstruction accuracy as long as it is used for a larger number of trials such that $n|\alpha|^2$ remains fixed.

\section{Discussions}
\label{sec:discussion}
In this section we first comment on two salient observations: the superiority of heterodyne measurements in reconstructing Gaussian devices as seen in Sec.\ref{sec:Scalability with the number of modes} and the utility of the reconstruction method beyond the Gaussian regime.

\subsection{Superiority of heterodyne measurements}
The results of the simulation in Sec.~\ref{sec:Scalability with the number of modes} demonstrate the superiority of heterodyne measurements for reconstructing Gaussian processes but not for linear processes. We now consider why this might be the case. 

We suggest that the superiority of heterodyne reconstruction is explained by the nature of the quantum state on which quadrature measurements are performed. 
In the context of quantum state reconstruction, it was shown in Ref.~\cite{Teo2017HomoHetero} that both heterodyne detection and homodyne detection perform equally well in reconstructing the first moments of any minimum-uncertainty state. However, if the state is not a minimum-uncertainty state, heterodyne reconstruction performs better. This result is directly relevant to our method, as our reconstruction is dependent on measuring the first moment of the output state of the unknown process.

While characterizing linear processes, the output state of our method is a minimum-uncertainty state, as linear processes map coherent probes to other coherent states. In contrast, the output state for a Gaussian process is not a minimum-uncertainty state, as Gaussian devices map coherent probes to other Gaussian states in general. This difference in the nature of the output states explains the superiority of heterodyne measurements in reconstructing Gaussian processes.
\subsection{Detecting non-Gaussian dynamics} Having shown that the proposed method can characterize Gaussian dynamics efficiently, we now show that it can also detect if the dynamics is beyond the Gaussian regime. To this end, we consider a simple model of the non-Gaussian dynamics in a single-mode cubic phase gate $e^{i\gamma\hat{X}^3}$~\cite{Gottesman2001}. Such systems lead to the transformation of the quadrature operators nonlinearly as~\cite{Budinger2024}
\begin{align}
    \hat{X}^{\text{out}} &= \hat{X}^{\text{in}} ,\nonumber \\
    \hat{P}^{\text{out}} &= \hat{P}^{\text{in}} + 3\gamma \left(\hat{X}^{\text{in}}\right)^2.
\end{align}
This nonlinearity shall be apparent as we characterize the system by the proposed method, as the supposed symplectic matrix elements will be found to have elements that depend explicitly on the probe used.  In this instance, when the system is probed with a real coherent state and the momentum quadrature is measured at the output, the ratio of the output quadrature to the input quadrature will explicitly depend on the input probe amplitude, $\it{viz.}$,
\begin{equation}
    \frac{\left\langle\hat{P}^{\text{out}}\right\rangle}{\left\langle\hat{X}^{\text{in}}\right\rangle} = 3\gamma \left\langle\hat{X}^{\text{in}}\right\rangle.
\end{equation}
Therefore, repeating said process with a different coherent probe will  indicate the presence nonlinearity beyond Gaussian dynamics.
\section{Conclusion}
\label{sec:conclusion}

We have presented a method for characterizing Gaussian quantum processes using only Gaussian resources. Our approach directly reconstructs the symplectic matrix that describes an arbitrary Gaussian process, requiring only $O(N^2)$ measurements for an $N$-mode system. This method is scalable, and is resilient to optical loss and phase-modulation errors. As this method uses only experimentally accessible resources such as coherent probes and quadrature detection methods, it is well-suited for implementation in current experimental setups.

Our results indicate that heterodyne detection outperforms homodyne detection in reconstructing Gaussian processes. However, for linear optical transformations, both detection methods yield similar results. This observation suggests that heterodyne measurements may prove to be advantageous in a wider variety of tasks in quantum information processing.

\section{Acknowledgements}
KVJ acknowledges support from Wheaton College via the G.~W. Aldeen Grant as well as the Junior Faculty Alumni Grant.

\bibliography{Gaussian_reconstruction}

\begin{appendix}
\section*{Appendix}

\section{Unitary and symplectic descriptions of linear optics}
\label{appendix:Relate_U_to_S}

Consider a $N-$mode linear optical device with $\hat{a}_i^\dagger\, (\hat{b}_i^\dagger)$ as the input (output) modes, where $i\in\{1,2,\ldots,N\}$. Let $\bf{\hat{a}^\dagger}, (\hat{b}^\dagger)$ be the vector of the input (output) mode operators. The input and output quadrature operators are defined from the mode creation and annihilation operators as 
\begin{align}
{\bf{\hat{X}}^{\text{in}}} &= \frac{1}{\sqrt{2}}\left({\bf{\hat{a}}} + {\bf{\hat{a}^\dagger}}\right) \nonumber \\
{\bf{\hat{P}}^{\text{in}}} &= \frac{1}{\sqrt{2}i}\left({\bf{\hat{a}}} - {\bf{\hat{a}^\dagger}}\right) \nonumber \\
{\bf{\hat{X}}^{\text{out}}} &= \frac{1}{\sqrt{2}}\left({\bf{\hat{b}}} + {\bf{\hat{b}^\dagger}}\right) \nonumber \\
{\bf{\hat{P}}^{\text{out}}} &= \frac{1}{\sqrt{2}i}\left({\bf{\hat{b}}} - {\bf{\hat{b}^\dagger}}\right)
\label{eq:quadinout}
\end{align}

The device implements the unitary transformation 
\begin{equation}
  {\bf{\hat{b}^\dagger}} = U\bf{\hat{a}^\dagger},
  \label{eq:Uab}
\end{equation}
where $U$ is an $N\times N$ unitary matrix. This relation can be alternatively expressed as 
\begin{equation}
  {\bf{\hat{b}}} = U^*\bf{\hat{a}},
\end{equation}
where the elements of $U^*$ are the complex conjugates of the corresponding elements of $U$.
Our aim is to describe the relation between this unitary matrix and the $2N\times 2N$ symplectic matrix $S$ that also represents the transformation of the quadrature operators as
\begin{equation}
  {\bf{\hat{r}}}^{\text{out}} = S{\bf{\hat{r}}}^{\text{in}}.
  \label{eq:rSr}
\end{equation}

To this end, let us first define the real and imaginary parts of the unitary matrix $U$ as 
\begin{align}
  U = \mathbb{Re}(U) + i \mathbb{Im}(U),
 \label{eq:ReIm_U}
\end{align}
where all the elements of the matrices $\mathbb{Re}(U)$ and $\mathbb{Im}(U)$ are real.

We now find the output quadrature operators in terms of the input quadrature operators. To this end, we substitute the mode evolution in \eqref{eq:Uab} into the definition of output quadrature operators in \eqref{eq:quadinout}. We obtain
\begin{align}
  {\bf{\hat{X}}}^{\text{out}} &= \frac{1}{\sqrt{2}}({\bf{\hat{b}}} + {\bf{\hat{b}^\dagger}}) \nonumber \\
  &= \frac{1}{\sqrt{2}}\left(U{\bf{\hat{a}^\dagger}} + U^*{\bf{\hat{a}}}\right) \nonumber \\
  &= \mathbb{Re}(U)\frac{({\bf{\hat{a}}} + {\bf{\hat{a}^\dagger}})}{\sqrt{2}}+\mathbb{Im}(U)\frac{({\bf{\hat{a}}} - {\bf{\hat{a}^\dagger}})}{\sqrt{2}i} \nonumber \\
  &= \mathbb{Re}(U){\bf{\hat{X}}^{\text{in}}} + \mathbb{Im}(U){\bf{\hat{P}}^{\text{in}}}.
  \label{eq:xout}
\end{align}
Similarly, we obtain
\begin{align}
  {\bf{\hat{P}}^{\text{out}}} &= \frac{1}{\sqrt{2}i}({\bf{\hat{b}}} - {\bf{\hat{b}^\dagger}}) \nonumber \\
  &= \frac{1}{\sqrt{2}i}\left(U{\bf{\hat{a}^\dagger}} - U^*{\bf{\hat{a}}}\right) \nonumber \\
  &= -\mathbb{Im}(U)\frac{({\bf{\hat{a}}} + {\bf{\hat{a}^\dagger}})}{\sqrt{2}}+\mathbb{Re}(U)\frac{({\bf{\hat{a}}} - {\bf{\hat{a}^\dagger}})}{\sqrt{2}i} \nonumber \\
  &= -\mathbb{Im}(U){\bf{\hat{X}}^{\text{in}}} + \mathbb{Re}(U){\bf{\hat{P}}^{\text{in}}}.
  \label{eq:pout}
\end{align}
Finally, we combine \eqref{eq:xout} and \eqref{eq:pout} in matrix form so as to explicitly find the form of the symplectic matrix $S$.
\begin{align}
  \begin{pmatrix}
     {\bf{\hat{X}}^{\text{out}}}\\
     {\bf{\hat{P}}^{\text{out}}}
  \end{pmatrix}
   = \begin{pmatrix}
     \mathbb{Re}(U) & \mathbb{Im}(U) \\
     -\mathbb{Im}(U) & \mathbb{Re}(U)
   \end{pmatrix}
   \begin{pmatrix}
     {\bf{\hat{X}}^{\text{in}}}\\
     {\bf{\hat{P}}^{\text{in}}}
  \end{pmatrix}.
  \label{eq:Sxp}
\end{align}
Comparing \eqref{eq:Sxp} with \eqref{eq:rSr}, we find 
\begin{equation}
  S = \begin{pmatrix}
     \mathbb{Re}(U) & \mathbb{Im}(U) \\
     -\mathbb{Im}(U) & \mathbb{Re}(U)
   \end{pmatrix}.
\end{equation}

\section{Uniform loss model in Gaussian devices}
  \label{appendix:loss_commutes}
Consider a Gaussian quantum device with uniform loss in all modes. This loss is modeled by a set of fictitious beam splitters with the same transmissivity $\eta$ in each mode. We demonstrate that this type of loss commutes with the device, meaning that all losses can be effectively moved to the beginning of the device. This was shown to be true in Ref.~\cite{Oszmaniec2018} for the restricted case of passive linear optics.

\begin{figure*}[]
  \centering
  \includegraphics[width=\textwidth]{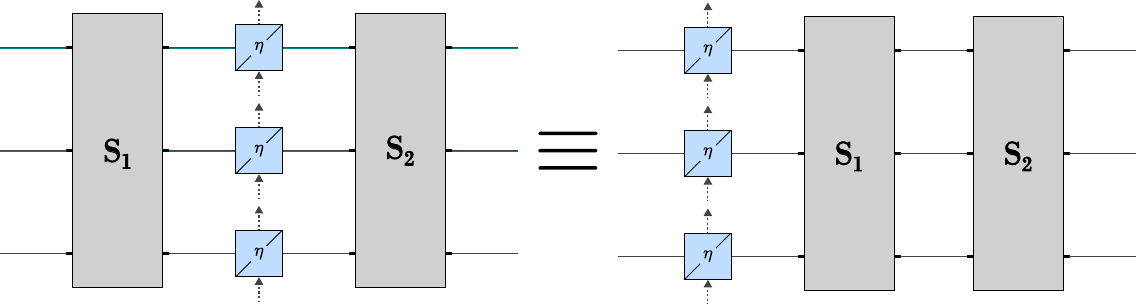}
  \caption{Schematic of Gaussian devices with uniform loss within and before the device. The transformation of the quadrature operators are identical in both cases.}
  \label{fig:gaussian_loss}
\end{figure*}

To this end, we consider a device with uniform loss within the device, and another device with uniform losses at the input. A schematic of these devices is shown in Fig.~\ref{fig:gaussian_loss}. We shall show that the transformations of the mode operators are identical in both these cases. 

If the device were to be lossless, then all beam splitters would have transmissivity $\eta = 1$, and the input quadrature operators transform as
 \begin{equation}
   {\bf{\hat{r}}} \rightarrow S_2S_1\bf{\hat{r}}.
 \end{equation}

Consider the scenario where uniform loss occurs within the device. In this case, the input quadrature operators of the device evolve as 
\begin{equation}
  {\bf{\hat{r}}} \rightarrow S_1{\bf{\hat{r}}}\rightarrow\sqrt{\eta}S_1{\bf{\hat{r}}}
  \rightarrow\sqrt{\eta}S_2S_1\bf{\hat{r}}.
  \label{eq:sandwiched_loss}
\end{equation}

In the alternate scenario where uniform loss occurs at the beginning of the device, the input quadrature operators evolve as
\begin{equation}
  {\bf{\hat{r}}} \rightarrow \sqrt{\eta}{\bf{\hat{r}}} \rightarrow\sqrt{\eta}S_1{\bf{\hat{r}}} 
\rightarrow\sqrt{\eta}S_2S_1{\bf{\hat{r}}}.
\label{eq:commuted_loss}
\end{equation}
\end{appendix}
The agreement of \eqref{eq:sandwiched_loss} with \eqref{eq:commuted_loss} shows that uniform loss within a Gaussian device can always be commuted to the beginning of the device. 

\end{document}